\documentclass[useAMS,usenatbib,usegraphicx]{mn2e}

\usepackage{amssymb}
\def\scla{8cm}
\def\sclauno{5.8cm}
\def\sclalita{4.2cm}
\def\sclalitata{4.105cm}

\title[High-resolution optical imaging of the core of the globular cluster M15 with FastCam]{High-resolution optical imaging of the core of the globular cluster M15 with FastCam}

\author[Anastasio D\'\i az-S\'anchez et al.]{
Anastasio D\'\i az-S\'anchez$^{1}$\thanks{E-mail: andiaz@upct.es}, Antonio P\'erez-Garrido$^{1}$, Isidro Vill\'o$^{1}$, Rafael Rebolo$^{2,3,4}$,
\newauthor 
Jorge A. P\'erez-Prieto$^{2,4}$, Alejandro Oscoz$^{2,4}$, Sergi R. Hildebrandt$^{2,4,5}$, Roberto L\'opez$^{2,4}$,  
\newauthor 
and Luis F. Rodr\'\i guez$^{2,4}$\\
$^{1}$Universidad Polit\'ecnica de Cartagena, Campus Muralla del Mar, Cartagena, Murcia E-30202, Spain\\
$^{2}$Instituto de Astrof\'\i sica de Canarias, C\ V\'\i a L\'actea s/n, E-38205, La Laguna, Spain\\
$^{3}$Consejo Superior de Investigaciones Cient\'\i ficas, Spain \\
$^{4}$Departamento de Astrof\'\i sica Universidad de La Laguna, La Laguna, Spain\\
$^{5}$California Institute of Technology, 1200 E. California Bvd. Pasadena 91125 USA\\
}

\begin{document}

\date{}


\pubyear{2010}

\maketitle

\label{firstpage}

\begin{abstract}
We present  high-resolution $I$-band imaging    of the core of the globular cluster M15  obtained at the  2.5 m Nordic Optical Telescope with FastCam, a  low readout noise L3CCD  based instrument. Short exposure times (30 ms) were used to record  200000 images (512 x 512 pixels each) over a period of 2 hours 43 min. The lucky imaging technique was then applied to generate a final image of the cluster centre with   FWHM$\sim 0''.1$ and $13''\times13''$ FoV. We obtained a  catalogue of   objects in this region with a limiting magnitude of  $I=19.5$. $I$-band photometry and astrometry are reported for 1181 stars. This is the deepest $I$-band observation of the  M15 core at this spatial resolution. Simulations show that crowding is limiting  the completeness  of the catalogue.  At shorter wavelengths, a similar  number of objects has been reported using   HST/WFPC observations  of  the same field. The cross-match with the available HST catalogues allowed us to produce colour-magnitude diagrams where we identify  new Blue Straggler star candidates and previously known stars of this class.
\end{abstract}


\section{Introduction}
The cores of globular clusters contain a very dense stellar population reaching up to 10$^6$ stars pc$^{-3}$. They are test laboratories for studying the dynamics in dense environments and provide crucial information on the evolution of the cluster. Encounters and direct collisions between core  cluster stars are highly probable leading to a variety of exotic species of stellar objects such as close interacting binaries, pulsars, x-ray sources and Blue Stragglers (BS). The globular cluster cores are also  potential hosts of intermediate mass black holes (\cite{Mc03}), which may be revealed by the dynamics of the stars in their very inner regions. Resolving the core stellar population is required to establish the existence and determine the  mass of any central compact object. Studies of the stellar populations in the globular cluster most inner regions have been conducted  in the optical using the  Hubble Space Telescope (see e.g. \cite{yanny94}, \cite{guha96}) and in the near infrared using  Adaptive Optics systems  in  ground based telescopes (see e.g. \cite{Davidge99}).

M15 is one of the oldest and  most massive globular clusters with an unusually compact core and a high central velocity dispersion (\cite{Mc03}). This cluster is particularly suitable to investigate the presence of a massive central compact object.  
High-resolution observations of the M15 core have already been  obtained  by \cite{yanny94} with the HST/PC camera with the filters F336W, F555W and
F785LP ($UVI$ bands) and by \cite{FP93} using F140W, F220W and F342W filters and the HST/FOC camera. These early data were taken with aberrated optics and it was difficult to perform accurate photometry of the core stars. After post-repair  HST  took images of the core  with the HST/WFPC2 with the filters F336W, F439W and F555W ($UBV$ bands) (\cite{guha96}), and more data in the $V$ band were reported by \cite{Mc03}. High resolution optical images using adaptive optics in the Canada-France-Hawaii Telescope have  been reported by \cite{GEB00}. \cite{piotto02} and \cite{VanderMarel02} also performed independent photometry of the data taken by \cite{guha96} and  in particular, the catalogue by \cite{piotto02} has been  used to study Blue Straggler stars by \cite{moretti08}. The inner region of  M15  has  been  studied more recently using deep FUV F140LP and NUV F220W images obtained with the HST/ACS camera (\cite{DK07}) which show  the presence of  horizontal branch stars, blue stragglers and white dwarfs. 

The development of fast readout very low noise  L3CCD  detectors has opened the possibility of obtaining high-resolution  images of the cores of globular clusters using ground based telescopes and the lucky imaging technique (\cite{LM06}). Here we  present lucky imaging observations in the I-band of the core of the globular cluster  M15 obtained with FastCam (\cite{Oscoz08}) at the 2.5m Nordic Optical Telescope. We report  $I$-band photometry for 1181 sources in a radius of $\sim$ 6.5 arcsec from the centre of  M15 and compare    with  available stellar catalogues from HST. We find that the lucky imaging technique provides reliable photometric measurements for stars in crowded fields. We take advantage of this technique to study known variable stars and to search for new objects in the core. 
The same  instrument was  used recently to obtain  high contrast optical imaging of substellar companions (\cite{LR10}) and combined with adaptive optics to produce high   precision astrometry of a  brown dwarf binary (\cite{FR10}).

\begin{figure*}
\begin{center}
\includegraphics[width=\sclauno]{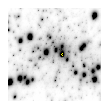}
\includegraphics[width=\sclauno]{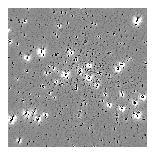}
\includegraphics[width=\sclauno]{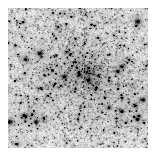}
\end{center}
\caption{
FastCam image of the central region of M15  (left panel) obtained using the lucky imaging technique, the yellow circle shows the location of the star that was used for image alignment. The scale plate is 0$''$.031 pixel$^{-1}$, and the field of view is $13''.16\times13''.16$. 
The same  image after convolution with  a ``mexican hat'' kernel to improve the detectability of point sources (central panel).
HST/WFPC2 (central chip, from Guhathakurta et al. (1996)) in the F555W filter (right panel). 
North is up and east to the left for all the images. They are displayed using a square-root stretch.}
\label{fig1}
\end{figure*}

\section{Observations and data reduction}

Observations were obtained on the night of 2008 July 25 with the FastCam instrument on the 2.5-m Nordic Optical Telescope using the $I$-band filter (820 nm)  which matches the Johnson-Bessell system. FastCam was equipped with a   512 x 512 pixels L3CCD from Andor Technology. The  optics was designed to provide a spatial scale of about  30 mas/pixel which was later determined accurately via comparison with available HST astrometry for M15 stars  and using  visual binaries from the Washington double star catalogue (\cite{WDS01}). 
We pointed the telescope to the centre of M15 (R.A.= $21^h 29^m 58^s.3$ and Dec= $12\degr 10' 1''$) and collected 200 series (``cubes'') of 1000 images each  with individual  exposures of 30 ms. These 200000 images were obtained during 2 hours 43 min of real observing time. The average seeing  during this observing period was $\sim$ 0.65 arcsec as measured in the final $I$-band image (combining all of the I-band images without applying 
the shift-and-add algorithm).

We made the raw data reduction in two steps. First, we selected the best 70  images of each cube (7\% of the total)  as follows. We defined a circular region on the first image of a cube with a 45 pixels diameter ( 1''.3), centred in the brightest star (ID 6290 in reference \cite{VanderMarel02}, it is marked with a yellow circle on the left panel of figure \ref{fig1}) near the centre of the field. This star was always the brightest available within this circular region.  We checked that in all the images the reference star was present
within this region. Then we identified the brightest pixel within this region for each image. We selected the 70  images with the highest values of the brightest pixel (i.e. 7 \% of the total). We determined the position of the brightest pixels and applied  a shift-and-add algorithm to these selected images and generated  one final image per cube. Our shift-and-add algorithm employs integral pixel shifts (sub-pixel shifts provide a slight improvement of order 5 \% of the FWHM of the final image with respect the case of integer shifts). In this way, we obtained  200 images  with an effective integration  time of 2.1 s each. This procedure was carried out for  different percent criteria and found that the best compromise between  sensitivity and spatial resolution was achieved by selecting a fraction of images  in the range 5-10\%, therefore we adopted the 7\% criterion.

As a  second step we took   a  reference image from these 200 images and defined a new circular region of 45 pixels of 
diameter centred in the previous bright star. Those images where the selected bright star was out of this 
region  were ruled out in order to avoid  too large shifts for the final combined image. Only two images were
ruled out of 200. Then we applied the shift-and-add algorithm again and obtained a final image of $512\times512$ pixels. Only in the central $422\times422$ pixels there is an effective integration time of 415.8 s, due to the maximum shift applied to  the images ($512-2\times 45$ pixels). This central field is shown on the left panel of figure \ref{fig1} and we will focus on this region to carry out the photometric work. On the right panel we display observations of the same field  with HST/WFPC2 (central chip) in the F555W (V) filter (\cite{guha96}).

\section{Analysis}

\begin{figure}
\includegraphics[width=\scla]{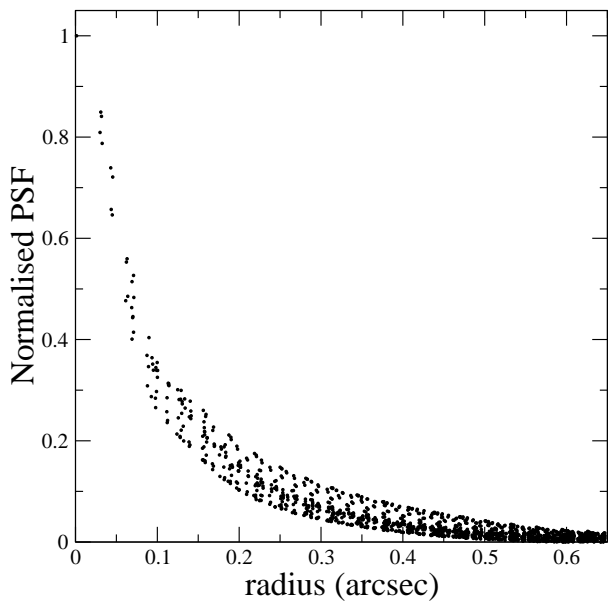}
\caption{
The average of the light profile of the 14 relatively isolated stars in our image. Here the light profile of each star is normalised and a PSF
template is formed with a shift-and-add algorithm at sub-pixel scales, each pixel of the PSF template is shown with a point in the figure.
}
\label{fig2}
\end{figure}

\begin{figure}
\includegraphics[width=\scla]{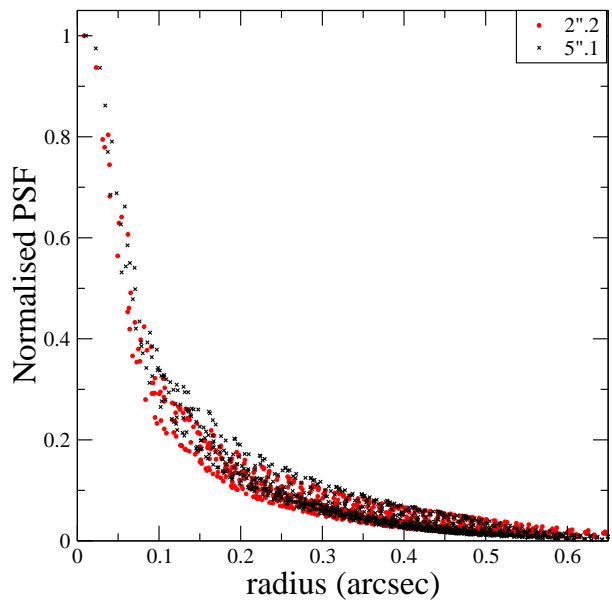}
\caption{
The normalised light profiles for two stars: the profile marked with red circles belongs to a star at a distance of
 2$''$.2 from the star that was used for image alignment and the profiles marked with crosses ($\times$) belongs to a star 
at a distance of 5$''$.1. Each point in the figures represents a pixel.
}
\label{fig2b}
\end{figure}

In Fig. \ref{fig2}, we show the average of the light profile of 14 relatively isolated stars in our 
image. Each stellar profile is normalised and a PSF template is built 
with a shift-and-add algorithm at sub-pixel scales, each pixel of the PSF template is shown with a point 
in the figure. We note the presence of a remarkable narrow core and an extended halo.
This is the characteristic profile of stars in images obtained using the lucky imaging technique as described in previous work (\cite{BW08}). 
The FWHM of the average normalised PSF in our final 
image is 0$''$.1 very close to the diffraction limit of 0$''$.084 of the telescope at the wavelength of the observations.
The normalised light profiles for two stars are shown in figure \ref{fig2b}, the profile marked with red circles belongs to a star at a distance of
 2$''$.2 from the the star that was used for image alignment and the profiles marked with crosses ($\times$) belongs to a star at a distance of 5$''$.1. There are slight differences between the two which may indicate a modest amount of anisoplanicity and that the dominant turbulent 
layer at the time of the observations was probably at very low altitude. It should be pointed out that the stars are not 
perfectly round. We do not have an obvious explanation for this, in particular it could be related to guiding issues which may be at play. We also note the relatively broad azimutal distribution of fluxes in the halo of the two stars which is caused by the presence of very close weaker contaminant stars.

\subsection{Detection of stars}
\label{secdetection}

 In order to carry out the detection of objects in our final image we applied  a derivative filter  to the whole image, we used the Laplacian of Gaussian filter, often called ``mexican hat'', which works efficiently in dense 
stellar fields (\cite{capella94}),
\begin{equation}
MH(x,y)=\frac{1}{2\pi\sigma_{\rm m}^4} \left( 2- \frac{x^2+y^2}{\sigma_{\rm m}^2}\right) \exp \left(-\frac{x^2+y^2}{2\sigma_{\rm m}^2}\right),
\end{equation}
where $\sigma_{\rm m}$ controls the width of the Gaussian and x, y determine the coordinates of the pixel. In the image convolved with the ``mexican hat'' (with $\sigma_{\rm m}=1$ pixel), 
the peaks are enhanced and the saddle points approach the background, so the detectability of  the sources  is definitely improved.

The local maxima are detected on the convolved image using a threshold, set at 3 times the standard deviation of the  background determined in the less crowded part of the image. In order to measure this background  we divided the image into squares  of $24\times24$ pixels and measured the standard deviation for each. The lowest of these values was adopted as the standard deviation ($\sigma$) of the less crowded part of the image.  We note that no region was  completely free of sources. We determined  the local maxima by  comparison of  each pixel with its 8 neighbours. In this way we generated a list of potential  objects  with their positions determined  as the centroid of a box of $3\times3$ pixels centred around each maximum. 
On the central  panel of figure \ref{fig1} we can see the resulting convolved image. 
1682 objects were detected in this image of $422\times422$ pixels ($13.16\times13.16$ arcsec$^2$). Only a very small fraction are false detections.

\subsection{Cross-match with HST/WFPC2 data}
\label{secHST}

The spatial scale of the image was obtained from a cross-match with the  available  HST/WFPC2 catalogue by \cite{VanderMarel02}  as described below. We first made a linear transformation of the catalogue into our list  and identified 762 common pairs with $V<19$ within a tolerance of 2 pixels of FastCam. We improved the number of matches among the two datasets using a second order transformation from the $x$-$y$ plane in our image to the $\Delta$R.A.-$\Delta$Dec. plane of the catalogue. The transformation equations  modelled a zero
point difference, a scale change, a rotation, a tilt, and second order distortion corrections. These transformations include 12 unknown
coefficients that are solved using a least squares method
\begin{equation}
\Delta{\rm R.A.} = a + b \Delta x + c \Delta y + d (\Delta x)^2 + e \Delta x \Delta y + f (\Delta y)^2,
\end{equation}
\begin{equation}
\Delta{\rm decl.} = a' + b' \Delta x + c' \Delta y + d' (\Delta x)^2 + e' \Delta x \Delta y + f' (\Delta y)^2,
\end{equation}
with $\Delta$R.A. = R.A. - R.A$._{\rm cent}$, $\Delta$decl. = decl. - decl.$_{\rm cent}$, $\Delta$x = x - x$_{\rm cent}$ 
and $\Delta$y = y - y$_{\rm cent}$,
where R.A.$_{\rm cent}$, decl.$_{\rm cent}$, x$_{\rm cent}$ and y$_{\rm cent}$ are taken from the object nearest to the centre in our image. 

Using  the initial list of 762 matched objects with $V<19$ we found the transformation coefficients using least squares. These coefficients were used to compute the positions of our objects in the $\Delta$R.A.-$\Delta$Dec. plane. For each object in our list, we determined distances to the objects in the HST catalogue, the minimum of these distances determines the correlation length for the object. We assume  that a FastCam object can be matched  with an HST object when the correlation length is smaller than $0.046$ arcsec ($1.5$ pixels). We removed from the original list used for the transformation  objects with correlation length larger than   $0.046$ arcsec and $V<19$ and performed an iterative process until we obtained a collection of FastCam objects which all have a correlation length smaller than such value. The final list of objects used to calculate the coefficients of the transformations have 770 objects with $V<19$ distributed uniformly  over the whole image. Then we used these transformations to cross-match the 1682 objects found by FastCam with objects in the catalogue of reference \cite{VanderMarel02} and finally selected those with  a correlation length smaller than $0.046$ arcsec, which corresponds to 1 pixel in the central chip of HST/WFPC2 \cite{VanderMarel02}. We obtained 1481 matched stars and 201 unmatched, we note  that  these unmatched  stars  include some potential false detections in the FastCam image.

In the catalogue given by \cite{VanderMarel02} and obtained from the images presented in \cite{guha96}, there are 3015 objects in 
the  field observed by FastCam. These authors used a combined F336W, F439W and F555W image ($UBV$ bands) to detect stars and added manually faint stars with peak brightness below the detection threshold of the DAOPHOT/FIND routine. From the same set of observations but using only the $B$ and $V$ images \cite{piotto02} detected 2221 objects in this field. We can increase the number of detections in our image by  decreasing the threshold used in subsection 3.1 to detect local maxima. For  a $2\sigma$ threshold we obtain 1911 objects matched with the catalogue by \cite{VanderMarel02} and 384 unmatched; and at $1\sigma$, we obtain 2275 matched  and 1452 unmatched objects. In this work we restrict the study to the 1682 objects detected at $3\sigma$ level. We removed the false detections among the 201 unmatched objects  by visual inspection and concluded that 46 out of these 201  were true stars, which included both new stars non reported by \cite{VanderMarel02} and stars with too large uncertainty in the measured positions to perform a reliable correlation.  For 41 of these stars the difference between the centroid position and the location of the maximum signal in the convolved image was more than 1 pixel. This difference is mainly caused by the presence of a brighter contaminating star in the vicinity. There are 5 unmatched stars. Finally, we verified that the transformation coefficients associated with tilt and second order distortion corrections were negligible, and therefore using a linear transformation was sufficient to determine a plate scale of $31.18 \pm 0.03$ mas/pixel for the FastCam image.

\subsection{Photometry}

In order to perform  photometry  we use the technique adopted by \cite{guha96} which incorporates a set of standard DAOPHOT routines. 
Firstly we construct an empirical PSF by iteration. In each iterative step we construct a PSF template of radius 22 pixels by averaging 14 bright and relatively isolated stars. Then we remove the neighbours of these stars using the PSF template from the previous iteration and calculate again the PSF template of these stars. 
In each iteration the quality of the PSF is improved because the neighbours are removed with increasing precision, convergence is reached after five iterations. From all the template functions  tested the penny2 function which vary linearly with the image position provides the best approximation to the actual PSF. The penny2 function is a Gaussian core with a Lorentzian profile, in which both components can be 
tilted. Then the final PSF template is fitted to all stars
detected on the FastCam image with the ALLSTAR routine in DAOPHOT. Due to the difficulty to find relatively isolated stars on the image, the PSF subtraction does not work perfectly and the stars are not subtracted completely. This is because the PSF used for subtraction of the flux of neighbour stars is determined locally using the brightest, nearest stars, but sometimes there are no bright stars which can help to define
the local PSF with very good S/N and therefore the correction may not be perfect.
This prevents to carry out direct PSF photometry in our image. Therefore, in order to obtain  accurate photometry we followed the hybrid method described by  \cite{guha96} which combines direct aperture photometry and PSF fitting. This method works as follows: we take  a star from our list of objects and subtract the PSF to each neighbouring star within a 44 pixel radius (two times the radius of the region used to calculate the PSF). Then aperture photometry is obtained for the star using a circular aperture of radius 2.5 pixels (0,078 arcsec)  with a local sky background measured in a surrounding annulus. This procedure was repeated for each  detected target and obtained instrumental magnitudes, I$_{fc}$.

\subsection{Photometric calibration}

\begin{figure}
\includegraphics[width=\scla]{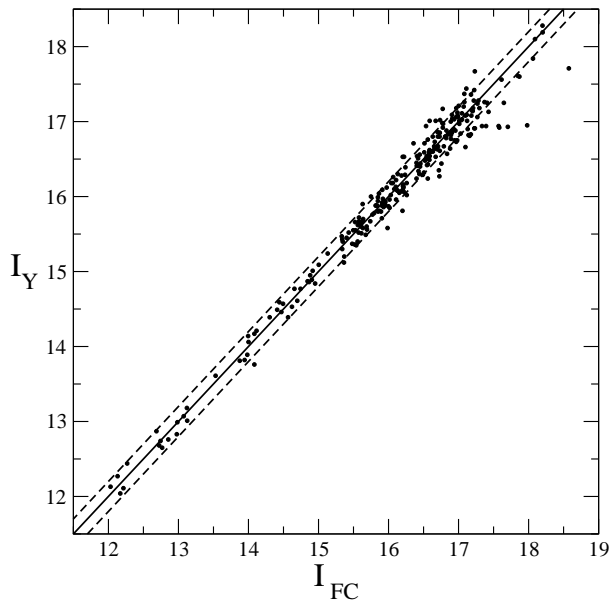}
\caption{Calibrated FastCam magnitude ($I_{FC}$) vs. Yanny et al.(1994) magnitude ($I_Y$) for the 269 stars in common. 
Dashed lines are the $\pm 2\sigma$ dispersion, stars within the region defined by dashed lines are chosen for the zeropint calibration.}
\label{fig3}
\end{figure}

The instrumental $I_{fc}$ magnitudes were converted to the Johnson $I$ standard system using  stars in  the catalogue by \cite{yanny94}. These  authors had  transformed their $F785LP$ instrumental magnitudes to Johnson $I$ system (\cite{J66}) and we directly adopted their $I$ magnitudes. We correlated with this catalogue  using the  algorithm described in subsection \ref{secHST} and  obtained 269 stars in common. The zeropoint of the calibration was obtained via an iterative process using only 219 stars in the magnitude range 12-18. The transformation was found to be linear and no colour term was found to be significant (figure \ref{fig3}).

\begin{figure}
\includegraphics[width=\scla]{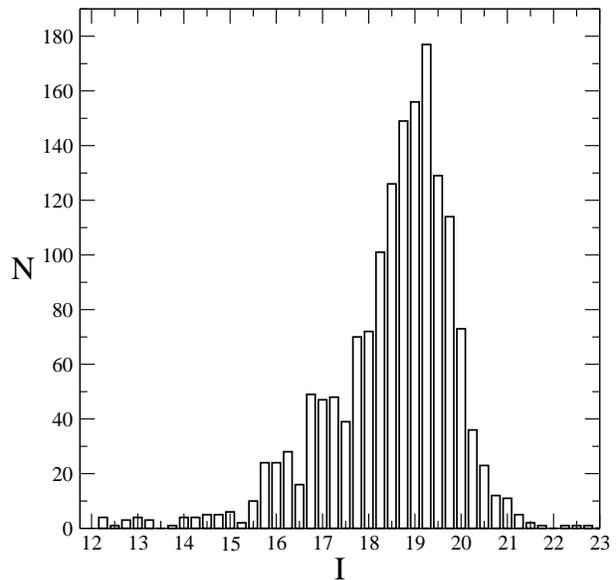}
\caption{Number of stars found with FastCam in 0.25 mag bins as a function of the $I$ magnitude.}
\label{fig4}
\end{figure}

The number of FastCam detected stars in 0.25 mag bins as a function of  $I$ magnitude is represented in figure \ref{fig4}.  We can see that this number increases until the 19-19.25 mag bin, and reach stars as faint as  $I\approx 21.5$. We removed  82 stars from our catalogue which had 
false photometry due to their proximity to much brighter stars.
Our catalogue provides $I$-band photometry of 1181 stars with $I\geq 19.5$, they are ordered from low to high $I$ magnitude and it is presented in its entirety in the electronic edition of MNRAS. 
For reference the first 10 entries are listed in table \ref{table1}. 

 \begin{table}
 \caption{FastCam $I$-band catalogue of the M15 core}
\label{table1}
\begin{center}
 \begin{tabular}{rrrrrr}
 \hline \hline
 ID & $\Delta{\rm R.A.}$ & $\Delta{\rm dec.}$ & $I$ & $V$&  ID$_{\rm VM}$ \\
 (1) & (2) & (3) & (4) & (5) & (6) \\
 \hline
    1 &    4.387 &   -0.053 &    12.03 &   13.369 &    4113 \\
    2 &    2.101 &   -2.216 &    12.13 &   13.696 &    5469 \\
    3 &    7.188 &   -6.430 &    12.17 &   13.605 &    5166 \\
    4 &   -4.717 &    2.919 &    12.21 &   13.352 &    6041 \\
    5 &   -0.110 &   -2.527 &    12.27 &   13.743 &    6290 \\
    6 &   -2.850 &   -1.718 &    12.69 &   13.977 &    6947 \\
    7 &   -5.211 &   -4.359 &    12.72 &   13.900 &    8408 \\
    8 &    5.711 &   -0.607 &    12.74 &   14.072 &    3891 \\
    9 &   -4.681 &   -6.160 &    12.77 &   13.870 &    8777 \\
   10 &    4.732 &   -6.416 &    12.85 &   13.952 &    5956 \\
 \hline \hline
 \end{tabular}
\end{center}
 \medskip {\em Notes.--} Table 1 is presented in its entirety in the electronic edition of MNRAS. The first 10 entries of the
$I$-band FastCam photometric catalogue described in the text are shown here for guidance regarding its form and content. The full catalogue contains 1181 stars. Col. (1) is the ID number of the star. Cols. (2) and (3) give coordinates (R.A., Dec.) for each star, measured in arcseconds with respect to the M15 cluster star AC 211 (which according to Kulkarni et al. (1990) has R.A.=$21^h 29^m 58^s.310$ and Dec= $12\degr 10' 02''.85$)). Col. (4)
gives the $I$ magnitude obtained with FastCam and cols. (5) and (6) the $V$ magnitude and the corresponding ID from the van der Marel et al. (2002) catalogue.
 \end{table}
 
\begin{figure}
\includegraphics[width=\scla]{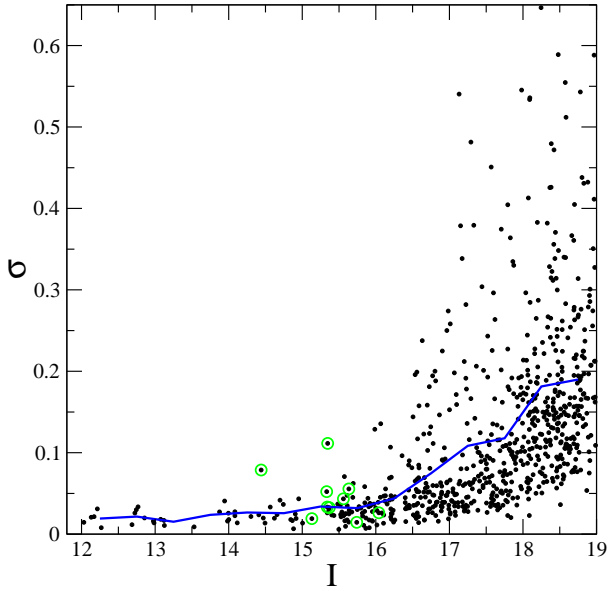}
\caption{Dispersion of the individual photometry of each star versus mean $I$ magnitude, for the 5 photometric measurements of combinations of 40 images as explained in the text. Variable stars from reference Dieball et al. (2007)
are denoted by green circles.}
\label{fig5}
\end{figure}

\subsection{Photometric accuracy}

In order to investigate the statistical errors of the photometry in the  final FastCam image, we bin the series of 200 images into combinations of 40 images each and perform the same photometric analysis described above. We therefore obtain the photometry for each star in our catalogue which is  bright enough to be detected in each of the 5 images. Figure \ref{fig5} shows the standard deviation of the 5 measurements as a function of stellar magnitude. In the magnitude range 12-15 the statistical errors are below 0.05 mag. In the plot we mark stars that are known to be variable (see section 4). In the final combined image we expect these statistical errors to decrease significantly and the final error to be dominated by the  systematic effects resulting from the variation of the PSF with position in the image and the residual contamination of poorly subtracted neighbouring stars.

\begin{figure}
\includegraphics[width=\scla]{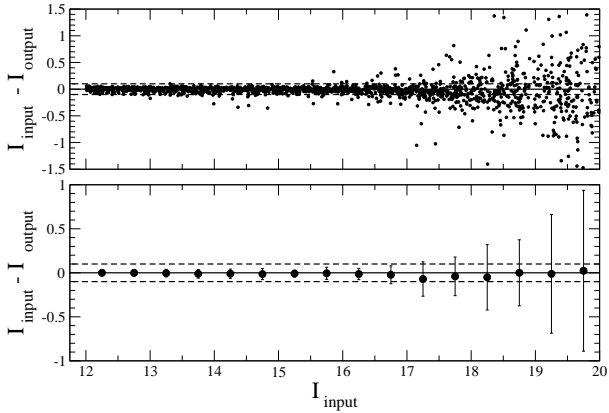}
\caption{Photometric errors  vs. $I$ magnitude as inferred from simulations. The top panel displays the difference between the input simulated magnitude and the output measured magnitude for each star as a function of  magnitude. On the bottom panel the mean value and the dispersion for each 0.5 mag bin (each simulated image) are shown.}
\label{fig6}
\end{figure}

\begin{figure}
\includegraphics[width=\scla]{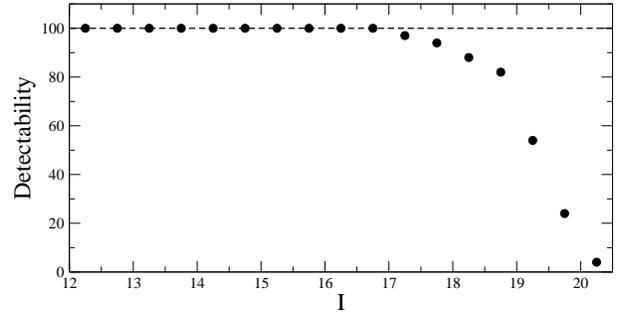}
\caption{Detectability of stars in our final image vs $I$ magnitude  as inferred from  the simulations. 100 stars were simulated 
at each 0.5 mag bin. The dashed line is the 100 \% detectability curve.}
\label{fig7}
\end{figure}

In order to study the impact of systematic errors we simulate stars using the previous PSF template obtained with the penny2 function which vary linearly with the position in the image. The stars are simulated in different positions on a $422\times422$ pixels image with initial zero background using standard DAOPHOT routines. We perform aperture photometry using a circular aperture of radius 2.5 pixels and find that the average  difference between the measured and  simulated magnitudes is less than  0.015 mag independent of the  simulated stars. Such  differences are smaller than the statistical error and could be  caused by a non-perfect adequacy of the PSF template due to  spatial variability.

\begin{figure}
\includegraphics[width=\scla]{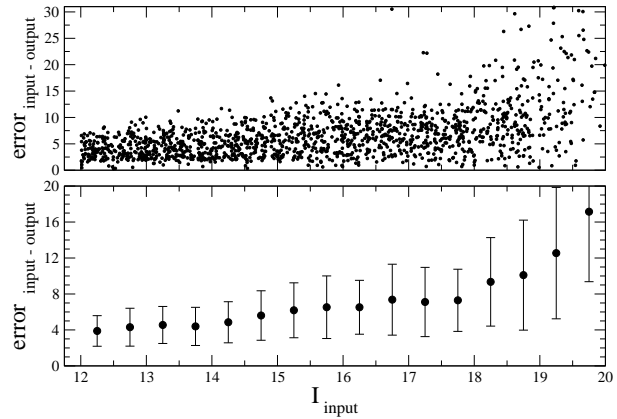}
\caption{Position errors (in milliarcseconds) versus $I$ magnitude of stars  as recovered from the simulations.
The top panel displays the difference in  position  between the input and the recovered  value from the simulated images. In the bottom  panel we show  mean values and  dispersions for stars in each 0.5 mag bin.}
\label{fig8}
\end{figure}

Errors associated to crowding were investigated simulating stars of a given magnitude with  a spatially constant PSF template.  We explored the magnitude  range I=12 to 20 mag simulating 100 stars  for each  step of 0.05 mag. In total, 16 images were generated each containing simulated stars of a given magnitude distributed uniformly on the final FastCam  image with a minimum imposed separation of 4 pixels from any two stars  (85\% of the stars detected by FastCam  have their nearest neighbours beyond 4 pixels and our detection algorithm only  consider stars with separation larger than 3 pixels). Stellar  magnitudes were then measured in theses images using  the hybrid method described above. In the top panel of figure \ref{fig6} we see the difference between the input and the output measured magnitudes for the simulated stars  as a function of  input  magnitude. At the bottom panel of this figure we show the mean value and the dispersion for each  magnitude bin. We conclude that there is no obvious bias  in our photometric measurements  associated to crowding. For the brightest objects the contribution of crowding to the photometric error appears to be less than 0.04 mag while for objects of magnitude $I=18$ is of order 0.2 mag. Therefore, in our final image  errors associated to crowding are  similar to the statistical errors for $I\lesssim 18$.

We also give in figure \ref{fig7} the number of  stars detected in the simulated images for each magnitude bin. For instance, in the 18.5-19 mag bin more than the 80\% of the objects  originally included in  the simulation  were detected by the  algorithm described in subsection 
\ref{secdetection}, and for the 19-19.5 mag bin more than 50\% were still detected.  To our knowledge this is the deepest and more complete high spatial resolution $I$-band observation of the M15 core reported so far.

 We also used the previous set of  simulations to investigate any possible bias in  our  astrometry measurements. The difference between the input and recovered  position for each artificial  star in the simulated images  is plotted at the top panel of figure \ref{fig8} in milliarcseconds. The positions are measured by the method described in subsection \ref{secdetection}. The mean value and the dispersion for each bin are shown at the bottom panel of figure \ref{fig8}.

\begin{figure}
\includegraphics[width=\scla]{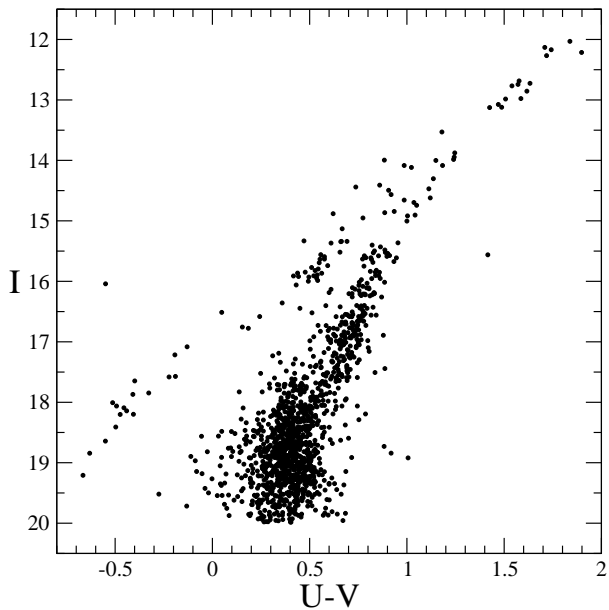}
\caption{
Colour-magnitude diagram $I$ vs. $U-V$. The $I$-band magnitude is from FastCam,
$U$ and $V$ magnitudes are from van der Marel et al 2002. There are 1312 objects correlated between
both catalogues with $I< 20$.
}
\label{fig9}
\end{figure}

\section{Stellar populations in the core of M15}

Figure \ref{fig9} displays the colour-magnitude diagram (CMD) $I$ vs. $U-V$ for the 1312 stars cross-matched  with the catalogue  of  \cite{VanderMarel02} and with $I< 20$. We note the main-sequence turnoff at $I\approx 19$. 
The CMD $V$ vs. $U-I$ shown in figure \ref{fig10} offers the advantage of a long colour baseline resulting  in a more clear separation of the various  types of stars. These are denoted in the figure with different symbols: blue stragglers ($\sq$, BS), bright red giant branch ($\nabla$, Bright RGB), (red) horizontal branch ($\times$, HB) and blue horizontal branch ($+$, BHB). The colour criteria used to assign  the various  stellar types  have been adopted according to \cite{yanny94} and are not strict. Number counts for the  various   types of stars found in the FastCam final full image are given in column 2 of table \ref{table2}.  For comparison with \cite{yanny94} we also list in columns 3 and 4  the number counts for  $r<5.6$ arcsec obtained in this work  and by these authors. We find no significant differences between the two. The CMD $I$ vs. $V-I$ is also shown in figure 
\ref{fig11}. Among the various types of stars that populate the core of M15 we have identified:

\begin{figure}
\includegraphics[width=\scla]{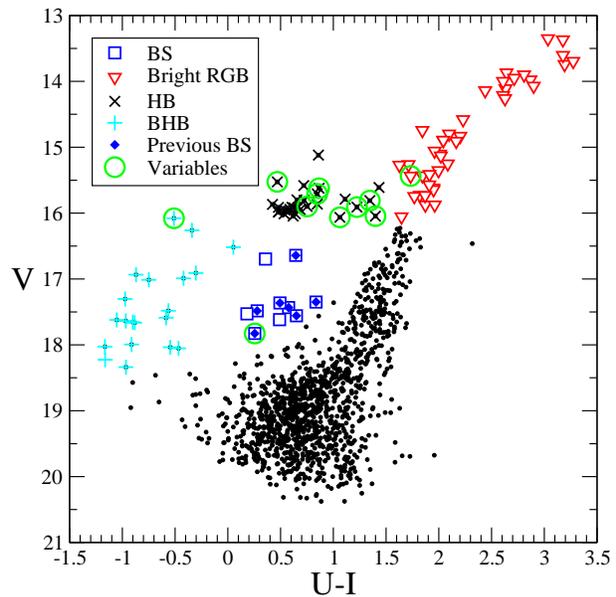}
\caption{
Colour-magnitude diagram $V$ vs. $U-I$ for the M15 core stars. $I$-band magnitudes are from FastCam (this work),
$U$ and $V$ magnitudes are from HST  (van der Marel et al. (2002)). Blue stragglers (BS),
bright red giant branch (RGB), (red) horizontal branch (HB) and blue horizontal
branch (BHB). Blue straggler stars given in previous references  are
marked with filled diamonds. Variable stars are taken from Dieball et al. (2007). AC 211 is the variable star in the BHB zone. 
}
\label{fig10}
\end{figure}

\begin{table}
\caption[]{Counts of various stellar types.}
\label{table2}
\centering
\begin{tabular}{c c c c}     % 4 columns
\hline\hline
 Stellar type & Counts & Counts   & Counts \\
 &  & ($r<5''.6$) & ($r<5''.6$) \\
 &  &  & (Yanny et al. (1994)) \\
\hline
Bright RGB & 40& 26 & 25 \\
HB & 29 &22 &18 \\
BS & 10 &7 & 7 \\
BHB & 20 & 12 & 12 \\
\hline \hline
\end{tabular}
\end{table}

\begin{figure}
\includegraphics[width=\scla]{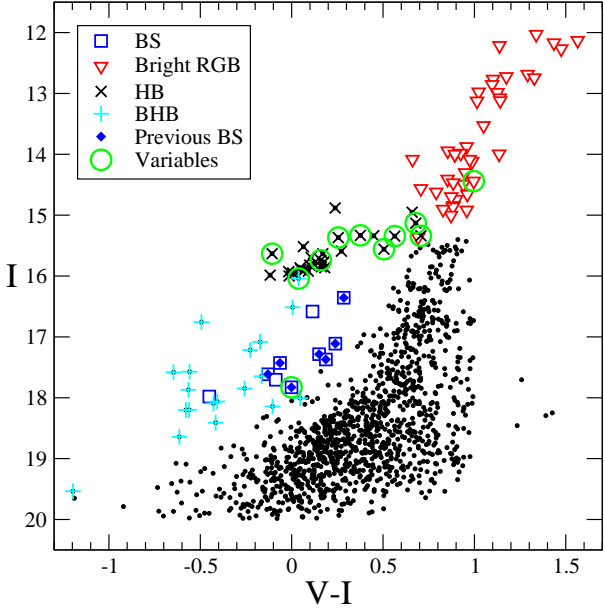}
\caption{
Colour-magnitude diagram $I$ vs. $V-I$ for the M15 core stars. $I$-band magnitudes are from FastCam (this work),
$V$ magnitudes are from HST (van der Marel et al. (2002)). Blue stragglers (BS),
bright red giant branch (RGB), (red) horizontal branch (HB) and blue horizontal
branch (BHB). Blue straggler stars given in previous references  are
marked with filled diamonds. Variable stars are taken from Dieball et al. (2007).
}
\label{fig11}
\end{figure}

\subsection {Blue straggler stars}

Blue straggler (BS) stars have  bluer colours and are brighter than main sequence turnoff  stars in globular clusters. The BS stars are frequently located in the densest regions of globular clusters where crowding makes  difficult their identification. Previous work on  M15 core stars  has produced BS candidates selected on the basis of CMDs in the the far-ultraviolet  \cite{FP93}, $I$ vs. $U-V$  \cite{yanny94} or  $V$ vs. $B-V$  (\cite{moretti08} using the data in  \cite{piotto02}). Previously known blue stragglers are marked in figure \ref{fig10} with filled diamonds.

In table \ref{table3} we list the BS stars found in the present work  and in \cite{yanny94,moretti08}.  Column one gives the ID number in our catalogue, the following three columns list the BS status according to  each  reference, and the last column gives some comments. The horizontal lines separate the  previously known BS stars (upper part),  new BS stars identified  in this work
(middle) and  BS star candidates in previous works which  are not confirmed as such here (bottom section).

\begin{figure}
\includegraphics[width=\scla]{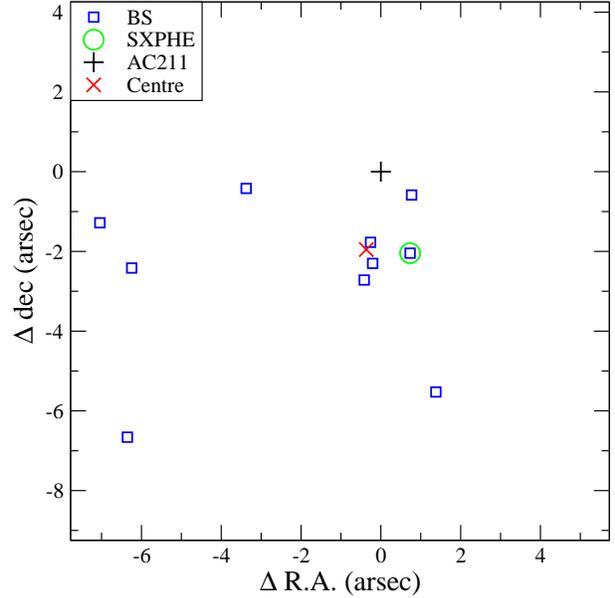}
\caption{
Location  of  blue straggler stars (BS) in the core of M15. The area showed is the field observed by  FastCam. The coordinates (R.A., Dec.) for each star are measured in arcseconds with respect to  star AC 211, marked with the $+$ symbol 
in the figure. The centre of the cluster given in van der Marel et al. (2002) is marked with a red cross. 
The SXPHE candidate is marked with a green circle. 
}
\label{fig12}
\end{figure}

We find 10 BS candidate stars (location plotted in Figure \ref{fig12}),  7 were previously known and 3 are new candidates. One  of the three new BS candidates  is located at less than 1 arcsec ($r< 1''$) from the cluster centre. Three out of the seven  previously known BS stars  also have a distance from the cluster centre of less  than 1.1 arcsec ($r< 1''.1$). 

We find in previous works 10 BS  candidates which  are not confirmed as such here. Only one of these 10 stars is listed in more than one reference (ID 145). For this object, we find a  $-1.41$ mag difference  between the $U$ magnitude by  \cite{yanny94} and the one by  \cite{VanderMarel02}. In the $V$-band the difference is $-0.8$ mag and comparing the $I$-band in \cite{yanny94} with ours  the difference  is 0.41 mag. This may suggest variability, but  the  star is not listed  as variable in \cite{DK07}. 

The other not confirmed BS stars are listed as candidates in only  one previous work, the three  found by  \cite{yanny94} are
at less than 1 arcsec ($r< 1''$) from the cluster centre where accurate photometry is more difficult to perform due to crowding.  The BS star candidate ID 369 is a variable star (ID V41 in \cite{DK07}), 
classified  as CV, it is likely  a SX Phoenicis (SXPHE) star. This object is not listed in the catalogue of \cite{yanny94}
possibly because of its variability.

\subsection {LMXBs}
\begin{figure}
\includegraphics[width=\sclalita]{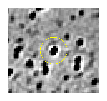}
\includegraphics[width=\sclalitata]{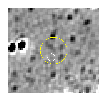}
\caption{
 FastCam $I$-band convolved images. Left panel: AC211 (M15 X-1), the yellow circle has a diameter of 0.5 arcsec and is centred in the AC211. Right panel: M15 X-2. The yellow circle is centred in the Chandra location of the X-ray detection and its diameter indicate the Chandra 0.5 arcsec error. The small white circle of diameter 0.14 arcsec marks the position of the FUV star associated to M15 X-2.
}
\label{fig13}
\end{figure}

Located in the inner region of M15 the V$\sim$ 15 star AC 211 is optically one of the brightest known low mas X-ray binary (LMXB) systems. It was identified by \cite{AF84} as the possible optical counterpart of the X-ray source 
4U 2127+119  and  \cite{CJ86} provided spectroscopic evidence for the classification as LMXB. The high optical to X-ray luminosity ratio suggests that a very luminous central X-ray source is hidden behind the accretion disk. A detailed analysis by \cite{IA93} revealed an orbital period of 17.1 hr. This object is clearly identified in our FastCam images (see the left panel of figure \ref{fig13}) as a source of  $I=16$ (ID 102). We have searched for photometric variability grouping our series of data  into 20 consecutive cubes (see the subsection on variability below). For each cube one final image was produced and photometry performed. The standard deviation of the series of 20 photometric measurements is $\sigma$=0.05 mag, consistent with the statistical error.   

A second luminous LMXB in the core of M15 was discovered by \cite{WA01} using the Chandra X-Ray Observatory. The Chandra observations resolved 4U 2127+119 into two X-ray sources, the previously known AC 211 and a second source  named as M15 X-2 (CXO J212958.1+121002). This new X-ray source is 2.5 times brighter  than AC 211 according to the Chandra counts rate and was associated with a $U=18.6$ mag star in the data from \cite{guha96} (star 590 in \cite{DM94}) located at $3''.3$ from the M15 centre. \cite{DK05} report time-resolved FUV photometry for the optical counterpart and argue that the FUV emission is dominated by an irradiated accretion disk around the neutron star primary. These authors concluded that M15 X-2 can be classified as an ultracompact X-ray binary. On the right panel of figure \ref{fig13} we show the location of the X-ray  source in our final FastCam image and indicate the position of the FUV star. We set a 3$\sigma$ upper limit to the I-band magnitude of this FUV source of $I=20.5$.

\begin{table}
\caption{Blue straggler candidates.}
\label{table3}
\centering
\begin{tabular}{c c c c l}     % 5 columns
\hline\hline
ID & This work & Yanny et al. & Moretti et al. & Comments \\
\hline
263 & Yes & Yes & Yes & \\
276 & Yes & Yes & Yes & \\
285 & Yes & Yes & Yes & \\
320 & Yes & Yes & Yes &  \\
239 & Yes & Yes & No & $r<1''$ \\
128 & Yes & No & Yes & $r<1''$ \\
369 & Yes & No & Yes & $r<1''.1$ \\
\hline
402 & Yes & No & No & $r<1''$  \\
156 & Yes & No & No &   \\
337 & Yes & No & No &  \\
\hline
145 & No & Yes & Yes & \\
103 & No & Yes & No & $r<1''$ \\
180 & No & Yes & No & $r<1''$ \\
220 & No & Yes & No & $r<1''$ \\
406 & No & No & Yes & \\
300 & No & No & Yes & \\
552 & No & No & Yes & \\
452 & No & No & Yes & \\
615 & No & No & Yes & \\
683 & No & No & Yes & \\
\hline
\hline
\end{tabular}
\end{table}

\subsection {RR Lyrae, Cepheids and cataclysmic variable stars}
RR Lyrae stars (IDs 43, 48, 57, 68, and 75, in our catalogue) and Cepheid stars (IDs 27, 44 and 46)  previously reported by  \cite{DK07} are  marked with  green circles in figure \ref{fig10}. We also detect  the cataclysmic variable  star (ID 369) which has colours suggestive of a SX Phoenicis (SXPHE)  star and note that the other known cataclysmic variable star (V39, \cite{DK07}) with a magnitude of $I\sim 21$ is very marginally detected (2$\sigma$ level) in our image. This object seems to have a rather blue colour when compared with previous  observations at shorter wavelengths  ($V=18.65$ and $U=19.08$).

\subsection{Variability: light curves of selected stars}

\begin{figure*}
\includegraphics[width=17cm]{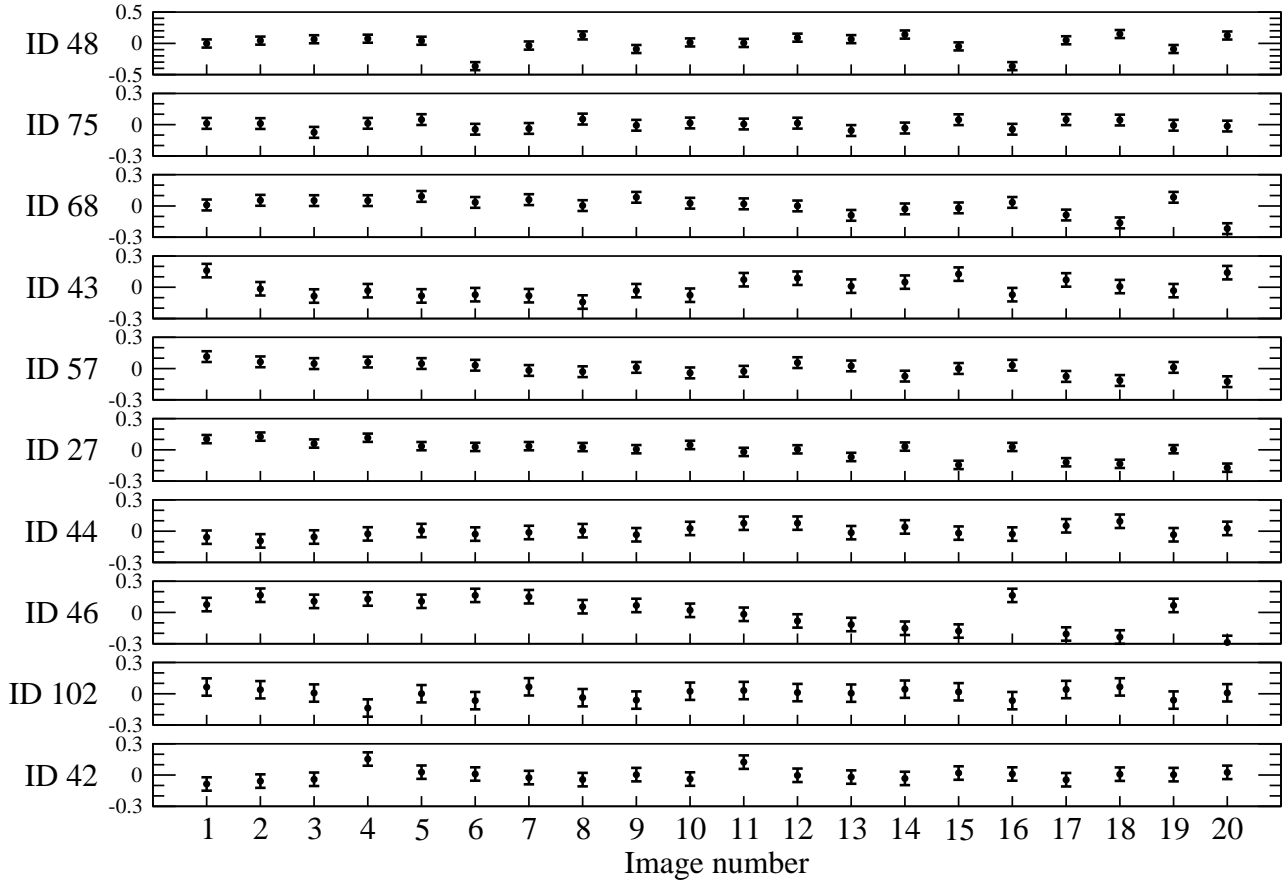}
\caption{
Mean-subtracted light curves (mag) for the variable sources previously identified as such in Dieball et al. (2007). There are 20 consecutive images which are represented in the $x$ axis.
Observations started at BJD-2454672.5215 and each image corresponds to a time interval of 8.1 min.
The RR Lyrae stars are IDs 48, 75, 68, 43 and 57 and they are identified as V9, V14, V23, V28 and V29 respectively in Dieball et al. (2007).
The Cepheid stars are IDs 27, 44 and 46 and they are identified as V10, V13 and V18 respectively in that reference. ID 102 is AC 211 and it is identified as V21 and ID 42 is identified as V24 in Dieball et al. (2007), where they found V24 was in the BHB zone but we find it is in the HB zone. 
Note that the scale of the upper curve is larger than the others. Errors are at 1-$\sigma$.
}
\label{fig14}
\end{figure*}

We have used our series of images to study the  short timescale  variability of stars of the previous types  with $I$ magnitude in the range  14.5-16 mag. For each set of  1000 images (each obtained over a period of 30 s real time)  we  generated a $lucky$ $image$. 200 such images were produced, each 10 consecutive images were then averaged in order to increase S/N and perform photometry on the  stars of interest. The total time span for the resulting series of 20 images is of 2 hours 43 min, each image corresponding to a time interval of 8.1 min. The series of photometric measurements obtained for a selection  of the  RR Lyrae, Cepheid, and LMXB stars in our image are shown in figure \ref{fig14}.
The majority of these stars show very little variability in the $I$-band. We note smooth trends in several of them  (ID 43, 44, 46) which are consistent with the known periods of variability. In some cases, we also detect changes among consecutive 8.1 min images which are clearly beyond the measurement errors and may be indicative of intrinsic short time variability, for instance due to pulses. Remarkably, the LMXB star AC 211 did not show evidence for any significant variability on these time scales.

\section{Conclusions}
Observations of the  M15 core with the 2.5-m Nordic Optical  Telescope using FastCam and the lucky imaging technique provided $I$-band images with   spatial resolution and sensitivity  close to those obtained by  HST in this region. We cross-match with the published HST/WFPC star catalogues and calibrate photometrically and astrometrically our image.  The number  of objects recovered in our final $13''\times13''$ FastCam image is comparable to that reported for the HST images.  A catalogue of 1181 stars is presented in this paper. Based on number counts the limiting magnitude of the catalogue is  $I\approx 19.5$. Errors in magnitudes and positions  are estimated from simulations. These simulations also indicate that crowding and  spatial resolution, more than sensitivity, limit  the completeness of the catalogue to about one magnitude brighter. Using a  CMD $V$ vs. $U-I$ we discuss the various stellar populations present in the M15 core. In particular, we identify a few new Blue Straggle star candidates and several new core stars which were not previously reported. We show that this imaging technique is particularly  useful
to investigate stellar populations in the core of globular clusters where the presence of bright stars may limit the use of more conventional techniques. Lucky imaging  observations of the core of M15 and other globular clusters undertaken with a baseline of several years may provide  proper motions of stars  in the very inner region of the clusters ($r<1''$) with the  precision required to constrain the properties of  intermediate-mass black holes.

\section*{Acknowledgments}

We thank the Nordic Optical Telescope staff and the Instrument Maintenance team of the IAC for their support during the observations. 
Based on observations made with the Nordic Optical Telescope operated on the island of La Palma by the NOTSA in the Spanish Observatorio 
del Roque de los Muchachos of the Instituto de Astrof\'\i sica de Canarias. This research has been supported by by Project No. 15345/PI/10 
from the Fundaci\'on S\'eneca and the Spanish Ministry of Economy and Competitiveness (MINECO) under the grant AYA2010-21308-C03-03.
This research has made use of the VizieR catalogue access tool, CDS, Strasbourg, France.
Based on observations made with the NASA/ESA Hubble Space Telescope, obtained from the data archive at the Space 
Telescope Science Institute. STScI is operated by the Association of Universities for Research in Astronomy, Inc. under 
NASA contract NAS 5-26555.

%\bsp

%\label{lastpage}

\end{document}